\documentclass[letterpaper]{article}

\usepackage[T1]{fontenc}

\usepackage{geometry}
\usepackage{setspace}

\usepackage[style = phys]{biblatex}
\usepackage{graphicx}
\usepackage{float}
\newfloat{scheme}{htbp}{los}
\floatname{scheme}{Scheme}
\floatname{chart}{Chart}
\newfloat{graph}{htbp}{loh}

\usepackage{chemformula} 
\usepackage[version = 4]{mhchem} 

\usepackage{authblk}
\author[1]{Francesco Pisani}
\author[2]{Baptiste Fix}
\author[1]{Thomas Krieguer}
\author[1]{Alexis Jouan}
\author[1]{Cheryl Feuillet-Palma}
\author[3]{Isabelle Sagnes}
\author[2]{Patrick Bouchon}
\author[1*]{Yanko Todorov}
\small{
\affil[1]{\small{Laboratoire de Physique et d'Étude des Matériaux, LPEM, UMR 8213, ESPCI Paris, Université PSL, CNRS, Sorbonne Université, F-75005 Paris, France}}
\affil[2]{DOTA, ONERA, Université Paris-Saclay, F-91123 Palaiseau, France}
\affil[3]{Centre de Nanosciences et de Nanotechnologies, CNRS – Université Paris-Sud-Paris-Saclay, 1 avenue Augustin Fresnel, 91120 Palaiseau, France}
}

\title{Engineering Nanophotonic Modes via the Radiation Continuum}
\date{*Email: yanko.todorov@espci.fr}

\begin{document}

\maketitle

\begin{abstract}
We demonstrate, experimentally and theoretically, a universal mechanism for combining nanophotonic modes relying on radiative-loss-mediated couplings. For the case of two modes this mechanism leads to a BIC-type phenomenon characterized by the emergence of a high-quality factor subradiant mode. This mode is experimentally observed in the mid-infrared range, in arrays of double-metal patch antennas where the radiation loss rates are controlled by the geometry of the system. As the mechanism described here is independent of the specific nature or number of the interacting modes, it can be used to combine physically different resonating structures, without requiring fine symmetry tuning or specific modal configurations, opening new opportunities for resonance-based nanophotonic devices.
\end{abstract}





\section{Introduction}
For many applications in nanophotonics it is desirable to have high quality factor resonators. The concept of photonic bound states in the continuum (BIC) has been emerging as an extremely powerful tool for designing and implementing coupled photonic modes with very long lifetimes \cite{PhysRevLett_2008, Bezus_2018, Hsu2016BIC, Koshelev2019BIC}. This concept has been realized across the electromagnetic spectrum with waveguides \cite{Bezus_2018}, photonic crystals \cite{Lee2012HighQResonances}, meta-surfaces \cite{Muhammad2022Radiationless}. BICs have been mainly induced by symmetry breaking of otherwise similar resonances \cite{PhysRevLett.121.193903}, topological means \cite{BykovBezusDoskolovich2020}, or by engineering a near field coupling \cite{Fix2017DoubleFabryPerot}.  Recently, the emergence of very high quality resonators have also been observed with interacting Fabry-Perot modes from Mid-IR and THz nanocavities \cite{Zhao_2020,Lackner2020NanoFabryPerot, Fix2017DoubleFabryPerot}.

\begin{figure}[ht]
    \centering
\includegraphics[width=0.7\textwidth]{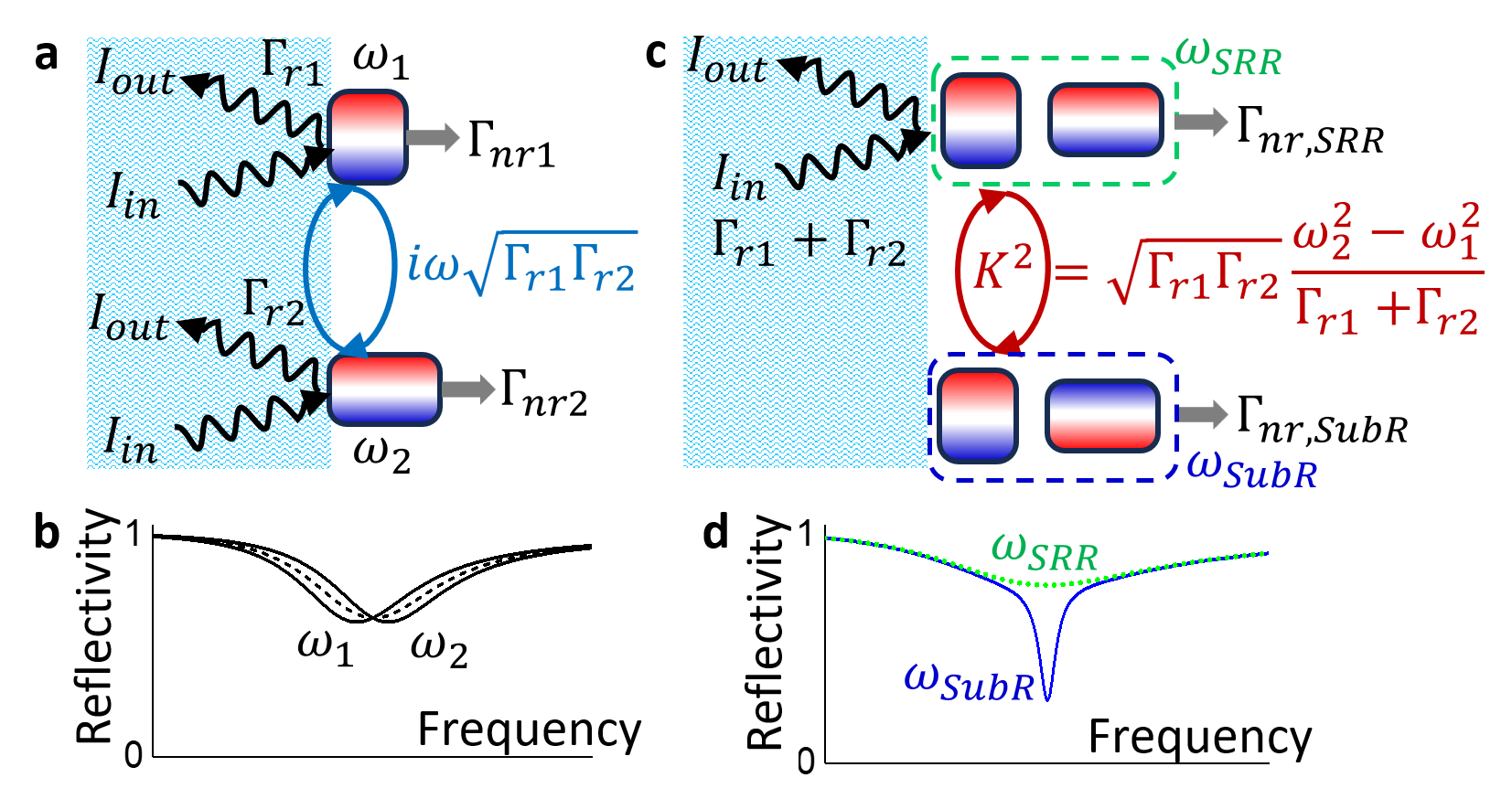}
    \caption{\textbf{a} Two resonators coupled to the free space (blue shaded area) by the coupling rates $\Gamma_{r1}$ and $\Gamma_{r2}$, and mutually  coupled with a term $i\omega\sqrt{\Gamma_{r1}\Gamma_{r2}}$.  \textbf{b} Reflectivity spectra from each resonator alone.  Dashed line: average of the two spectra. \textbf{c} After a linear transform, the system is mapped into a subradiant "SubR" mode coupled to a superradiant "SRR" mode with a coupling constant $K^2$. \textbf{d} The reflectivity spectrum of the combined system appears as a sharp resonance $\omega_{SubR}$ (full line) over a broad background centered at $\omega_{SRR}$ (dotted line).}
    \label{figI}
\end{figure}

Here, we describe and demonstrate experimentally a new and  general mechanism for recoupling the resonances from discrete and disparate systems. This mechanism does not depend on the specific nature of the electromagnetic modes, but only on their ability to couple with free space radiation. Namely, when all resonators are constrained to interact with the same radiation continuum, energy conservation imposes the emergence of mode-mode coupling terms expressed uniquely from the individual radiation loss rates. As a result, there is a strong renormalization of the resonant frequencies and resonators linewidth. In the case of two resonances, these couplings lead to a Friedrich–Wintgen BIC-type  phenomenon \cite{Hsu2016BIC} characterized by the emergence of a high-quality factor subradiant mode. Here we demonstrate this phenomena by employing an experimental system with controllable radiation loss, and we find excellent agreement between measurements and theory. Our theory is very general and, crucially, it can incorporate arbitrarily high radiatiave and non-radiative loss rates, providing a generalization of the first order temporal coupled-mode theory (CMT) \cite{Fan:03}. In this framework the physics of the quasi-BICs described here is linked to non-Hermitian description of nano-photonic modes \cite{Lalanne_2020, Roskos_2023, Wu2025}. 

\section{Model outline}

Considering a general multi-mode setup, in the framework of CMT \cite{Haus1984WavesFieldsOptoelectronics, Fan:03} each mode $m$ with a frequency $\omega_m$ is assigned a non-radiation loss rate $\Gamma_{nr m}$, accounting for internal loss, and a radiation loss rate $\Gamma_{r m}$ quantifying its coupling with free space. CMT is a perturbative approach based on the rotating wave approximation (RWA), yielding first order differential equation for the mode amplitude\cite{Haus1984WavesFieldsOptoelectronics, Fan:03}. Our approach is  based on a multi-mode and non-perturbative generalization of CMT previously described in Ref.\cite{Krieguer_2025}. We consider an input field $I_{in}$ exciting the resonators and a single reflection channel, with an outcoming amplitude $I_{out}$ (Fig.~\ref{figI}\textbf{a}). The resonator field is described by two quadratures, $D_m$ and $A_m$, that play the role for respectively the electric and magnetic field. They are normalized such as the electromagnetic energy density per mode is $w_m = (D_m^2+A_m^2)/2\varepsilon_0$, with $\varepsilon_0$ the vacuum permittivity. The quadrature $D_m$ is linked to $A_m$ through  $dD_m/dt = -\omega_m A_m$, while  $A_m$ is linked to the incident field by the equation:

\begin{eqnarray}
\frac{dA_m}{dt} &=&  \omega_{m} D_m - (\Gamma_{nrm}+\Gamma_{rm}) A_m \nonumber \\
 &-& \sum_{m' \neq m} \sqrt{\Gamma_{rm} \Gamma_{rm'}} A_{m'} +   2\sqrt{\Gamma_{rm}} I_{in}. \label{eqAm}
\end{eqnarray}

The output field is provided by $ I_{out} = -I_{in} + \sum_m \sqrt{\Gamma_{r m}} A_m$. Thus, in  \eqref{eqAm} the mode $A_m$ is coupled to all other modes of the system through coupling coefficients $\sqrt{\Gamma_{r m} \Gamma_{r m'}}$. Similar coupling terms can be deduced from the first order CMT \cite{Fan:03} where a single mode waveguide is considered instead of free space. As in Ref. \cite{Krieguer_2025}, we introduced the mutual couplings in order to satisfy the energy conservation condition:

\begin{equation}
\sum_m \frac{dw_m}{dt}  = \frac{1}{\epsilon_0} ( I_{in}^2 - I_{out}^2 - \sum_m \Gamma_{nr  m} A_m^2 ),
\end{equation}

 stating that the energy stored in the resonators is increased through the input Poynting flux $I_{in}^2$, and decreases through the output flux $I_{in}^2$ as well as through non-radiation internal losses of the resonators (terms $\Gamma_{nr  m} A_m^2$). We will see that the mutual coupling terms in \eqref{eqAm}  have  real experimental consequences, and modify profoundly the reflectivity spectrum of the system. 

The full details of the theory are described in \textcolor{blue}{Supporting Information}. We consider permanent harmonic excitation regime $I_{in} \propto \exp{(i\omega t)}$. First, considering a single resonant mode $m=1$, we can express the reflectivity $R_m(\omega) =|I_{out}/I_{in}|^2 = |1-i\omega\Gamma_{rm}/\Delta_m(\omega)|^2$. Here $ \Delta_{m} (\omega) = \omega_{m}^2 - \omega^2 + i\omega (\Gamma_{rm}+\Gamma_{nrm})$ is the characteristic functions of a harmonic oscillator damped by a total loss rate $\Gamma_{rm}+\Gamma_{nrm}$. This equation reproduces the result of CMT when RWA is applied. A second case of interest, relevant for BIC as well as prior experimental work \cite{Lackner2020NanoFabryPerot, Fix2017DoubleFabryPerot}, is that of two coupled modes $m=1,2$. The equation set \eqref{eqAm} then provides a system of two linear equations for the amplitudes $A_1$ and $A_2$ (\textcolor{blue}{Supporting Information}). As shown in the diagram of Fig.~\ref{figI}\textbf{a}, both modes interact with free space with rates $\Gamma_{r1}$ and $\Gamma_{r2}$, while they are mutually coupled with a term $i\omega \sqrt{\Gamma_{r1} \Gamma_{r2}}$. The individual reflectivity spectra of the modes $R_{m=1} (\omega)$ and $R_{m=2} (\omega)$ are illustrated in Fig.~\ref{figI}\textbf{b}. If the coupling term $i\omega \sqrt{\Gamma_{r1} \Gamma_{r2}}$ was ignored, then the combined spectrum from both $m=1$ and $m=2$ would appear as the mean value $(R_{m=1} (\omega)+R_{m=2} (\omega))/2$ provided by the dashed line in Fig.~\ref{figI}\textbf{b}. 

As shown in \textcolor{blue}{Supporting Information} we can perform a linear transformation of the system, $(A_1, A_2) \rightarrow (\tilde{A}_{1}, \tilde{A}_{2})$, which leads to the diagram of Fig.\ref{figI}\textbf{c}. Now, there is a sub-radiant mode, with a frequency $\omega_{SubR}$ that is completely decoupled from free space, and a superradiant mode, $\omega_{SRR}$ with a radiation loss rate $\Gamma_{r1}+\Gamma_{r2}$. The modes $SRR$ and $SubR$ are still coupled through the real coupling term $K^2 =\sqrt{\Gamma_{r1}\Gamma_{r2}/(\Gamma_{r1}+\Gamma_{r2})} (\omega_1^2-\omega_2^2)$. This residual coupling allows to funnel energy from free space to the $SubR$ mode via the $SRR$ mode. In a sense, the $SRR$ mode acts as an "antenna" for the $SubR$ mode. With the assumption $\Gamma_{nr1} \approx \Gamma_{nr2} \approx  \Gamma_{nr}$ we can express the reflectivity of the combined system as: $R(\omega) = |1-i\omega (\Gamma_{r1}+\Gamma_{r2})/(\Delta_{SRR}(\omega)-K^4/\Delta_{SubR}(\omega))|^2$. Here $\Delta_{SRR} (\omega) = \omega_{SRR}^2 - \omega^2 + i\omega (\Gamma_{nr} + \Gamma_{r1} + \Gamma_{r2})$ and  $\Delta_{SubR} (\omega) =  \omega_{SubR}^2 - \omega^2 + i\omega \Gamma_{nr}$. This equation shows that the broadening of the  $SubR$ mode is mainly provided by the non-radiation contribution $\Gamma_{nr}$, in addition to a small contribution $\Gamma_K \approx K^4/(\omega_{SubR}^2(\Gamma_{nr} + \Gamma_{r1} + \Gamma_{r2}))$ which is derived in \textcolor{blue}{Supporting Information}.  Thus, as illustrated in Fig.~\ref{figI}\textbf{d}, the spectrum of the combined system (blue curve) appears as a sharp resonance $SubR$ sitting on a broad background. The background (green dotted line) is identified with the sole contribution from $SRR$, setting $K=0$ in the full reflectivity expression. 

Thus, Fig.~\ref{figI} provides a guide for the experimental test of the model. Namely, the observation of the spectrum displayed in Fig.~\ref{figI}\textbf{d} will constitute a validation of the relevance for the mutual coupling $\sqrt{\Gamma_{r1}\Gamma_{r2}}$ between modes mediated by free space, and will simultaneously provide a route to engineering BIC-type resonances with high quality factors. 

\begin{figure}[h]
    \centering
\includegraphics[width=0.8\textwidth]{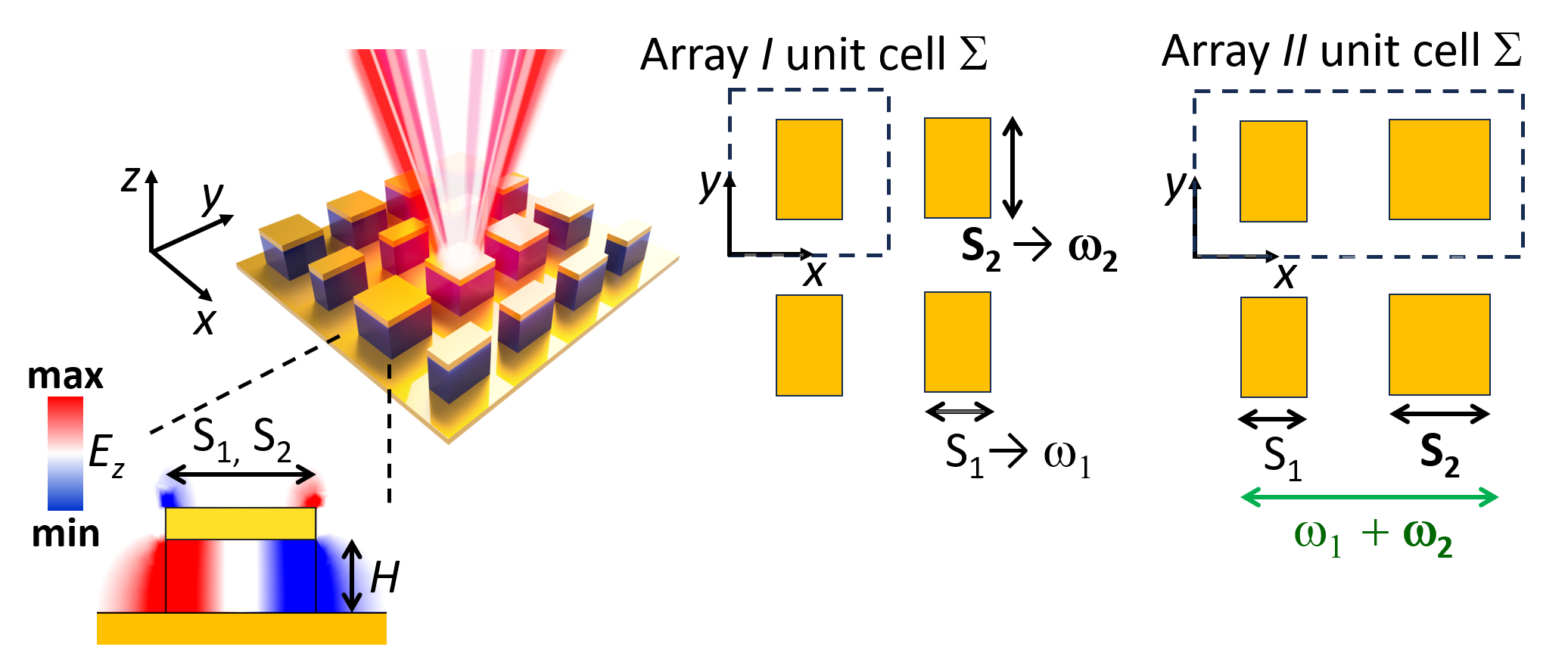}
    \caption{Arrays of metal-dielectric-metal resonators prepared for experiments, with their relevant dimensions. The $x-$ and $y-$ axis of the coordinate frame provide the polarizations of the incident light. 
    The blown-up section illustrates the electric field distribution for the  fundamental standing wave mode supported by the patch antennas. On the right, we show the geometry of array I used to characterize individual modes $\omega_{1,2}$ and array II used to prove their combined response. Variables related to the mode $2$ are underlined in bold text. }
    \label{figII}
\end{figure}

\section{Samples and Measurements}

\begin{figure*}[h]
   \centering
\includegraphics[width=0.95\textwidth]{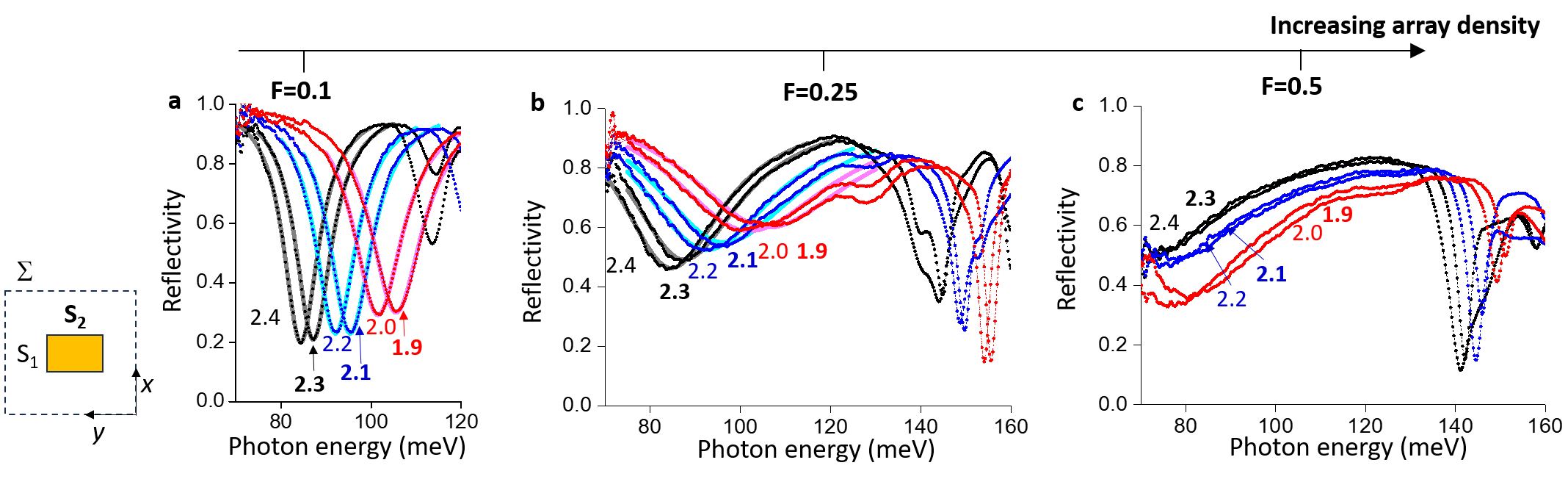}    
   \caption{Single mode spectroscopy. \textbf{a-c} Polarized
   reflectivity spectra of arrays with rectangular patches $S_2 \times \mathbf{S_1}$ (left inset), and increasing array filling factors $F$. Black, blue and red dotted curves correspond to measurements, with the values $S_2$ ($y$-polarization) and $\mathbf{S_1}$ ($x$-polarization) indicated for each pair of curves. The gray, cyan and magenta full lines correspond to fits with single mode reflectivity expressions.}
    \label{figIIIi}
\end{figure*}

\begin{figure*}[h]
   \centering
\includegraphics[width=0.95\textwidth]{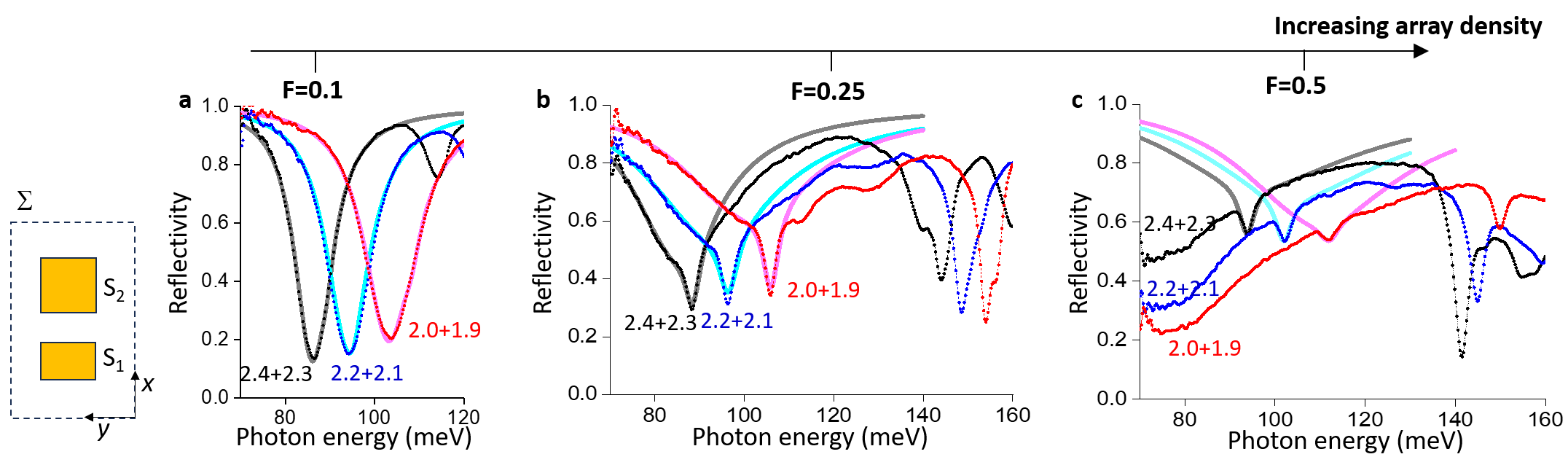}   
   \caption{Coupled mode spectroscopy. \textbf{a-c} $x$ -polarized reflectivity spectra from the arrays with combined $S_1+S_2$ patches (left inset) and increasing array filling factors $F$. Dotted curves are experiments, and full lines are reflectivity fits from our model.}
    \label{figIIIii}
\end{figure*}

To test our model, we use arrays of metal-dielectric-metal patch antennas supporting standing wave modes \cite{todorov_optical_2010} as described in Fig.~\ref{figII}. We use $S_1 \times S_2$ rectangular patches which support standing wave resonances with frequencies $\omega_1 = \pi c/n S_1$ and $\omega_2 = \pi c/n S_2$, vibrating along the respective patch width $S_{1,2}$ \cite{todorov_optical_2010}. The radiation loss rates of both modes depends on the resonator and array geometry through $\Gamma_r= cH/2n\Sigma$  where $c$ is the speed of light, $H$ is the thickness of the dielectric layer (Fig. \ref{figII}), $\Sigma$ is the array unit cell surface, and $n$ is the effective index of the resonant mode \cite{Krieguer_2025,Rodriguez_22}.

We prepared two types of arrays designed to test the two modes $\omega_1$ and $\omega_2$ separately, "array I", as well as the combined response when both modes are excited, "array II" (Fig.~\ref{figII}). Array I is composed solely of rectangular $S_1 \times S_2$ patches, 
(Fig.\ref{figII}, middle). This array  exhibits single resonant frequency $\omega_1$ or $\omega_2$ depending on whether it is illuminated with light polarized along the x- or y-axis. Its array unit cell is $\Sigma=P^2$ where $P$ is the array period that is identical along the  $x$- and $y$-axis.
Array II is a  combination of square $S_2 \times S_2$ and rectangular $S_1 \times S_2$ patches within the same unit cell $\Sigma=2P^2$ (Fig.\ref{figII}, right). An $x$-polarized light excites both modes  $\omega_1$ and $\omega_2$, allowing to test the model.

For each type of array, we have  $S_2 = S_1+100$ $nm$ and we varied  $S_1 =1.9$ $\mu m$, $2.1$ $\mu m$ and $2.3$ $\mu m$. We also varied the array periodicity in a way that the filling factor $F = \Sigma_{patch}/\Sigma$ is constant for both types of arrays, with either $\Sigma_{patch} = S_1S_2$ (array I) or $\Sigma_{patch} = (S_1+S_2)S_2$ (array II). For each type we fabricated three different families with $F=0.1$, $0.25$ and $0.5$ thus increasing progressively the radiation loss because of the dependence $\Gamma_r \propto 1/\Sigma$. The dielectric slab of the patch antennas is Gallium Arsenide (GaAs) with three different thicknesses: $H=0.15$ $\mu m$,  $0.3$ $\mu m$, and  $1.5$ $\mu m$.  Here we report the results for the $H=1.5$ $\mu m$ samples, where the radiation loss rate is sufficiently high to observe BIC;  the results with thinner samples are reported in  \textcolor{blue}{Supporting Information}. All arrays have a total area of $100 \times 100$ $\mu$m$^2$. The reflectivity of each array is measured with a micro-FTIR system, using a Cassegrin objective, with an aperture of $\sim50\times50$ $\mu$m$^2$ for both $x$- or $y$-polarized light. Each spectrum was normalized using an area of the sample in which no patches were present (gold reference). 

Figs. \ref{figIIIi} \textbf{a-c} report spectra from arrays I with increasing filling factor $F$. In the spectra Fig.~\ref{figIIIi} \textbf{a},\textbf{b} one can clearly see the reflectivity dips that correspond to the fundamental mode, either $\omega_1$ or $\omega_2$ depending on the polarization. These dips are fitted with the single mode expression $R_{m=1,2} (\omega)$, which allows extracting the parameters $\omega_m$, $\Gamma_{rm}$ and $\Gamma_{nrm}$ for each mode. All frequencies are well fitted with the expression $\omega_{1,2} = \pi c/n S_{1,2}$ with an effective index $n=3$, slightly different from bulk GaAs ($n=3.3$) as expected \cite{todorov_optical_2010}. We observe typical values for the non-radiation loss $\Gamma_{nrm} = 3-4$ $meV$ for all filling factors. For the radiation loss, we obtain respectively  $\Gamma_{rm} = 7-10$ $meV$ ($F=0.1$) and  $\Gamma_{rm} = 23-40$ $meV$ ($F=0.25$), confirming the strong increase of $\Gamma_{rm}$ with $1/\Sigma$. Actually, this dependence can also be derived from our model as explained in \textcolor{blue}{Supporting Information}. Higher order modes are also visible in the spectra. These include quadrupole modes at an energy roughly $\sqrt2$ times  the fundamental, and a second order at twice the fundamental. While dark  at normal incidence, these modes become visible \cite{todorov_optical_2010} because of  the large numerical aperture of the micro-FTIR objective (NA=0.4). As seen from Fig.~\ref{figIIIi}\textbf{c} the first order resonance for the densest $F=0.5$ arrays is completely damped by radiation loss, rendering the fitting impossible; nevertheless, we can still clearly see the higher modes. Since their energies are very similar with those for the $F=0.25$ structures, we conclude that the clear absence of the fundamental mode is indeed due to strong broadening, rather than other reasons. The baseline in the reflectivity signal for the $F=0.5$ arrays may also be attributed to increased scattering by fabrication defects around the patches (additional details are provided in \textcolor{blue}{Supporting Information}).

 The reflectivity spectra from arrays II are shown in  Fig.~\ref{figIIIii}\textbf{a,b,c}. For the $F=0.1$ arrays (Fig. ~\ref{figIIIii}\textbf{a}), the combined spectra are typically larger than the individual resonances (see Fig.~\ref{figIIIi}\textbf{a}), with a lineshape that is roughly an average of two line-shapes of the individual modes. Here, the radiation loss rates $\Gamma_{r1}$,$\Gamma_{r2}$ are too small to induce coupling between the modes. Similar behavior is observed in the thinner samples (\textcolor{blue}{Supporting Information}). Remarkably, for higher array densities, $F=0.25$ (Fig.~\ref{figIIIii}\textbf{b}) and $F=0.5$ (Fig. \ref{figIIIii}\textbf{c}), where the radiation loss rates are high, we do observe a narrow resonance that appears in the combined spectra, sitting on a broad background, very similar to the prediction of Fig.~\ref{figI}\textbf{d}. Spectra are fitted with the reflectivity expression in the basis of $SubR$ and $SRR$ modes  $R(\omega) = |1-i\omega \widetilde{\Gamma}_{r}/(\Delta_{SRR}(\omega)-K^2/\Delta_{SubR}(\omega))|^2$. The fitting parameters are the frequencies of the two modes $\omega_{SRR}$, $\omega_{SubR}$, the overall radiation loss $\widetilde{\Gamma_{r}}$, the non radiation loss $\Gamma_{nr}$, and the coupling constant $K^2$. Excellent match with the data are obtained for the $F=0.1$ and $F=0.25$  arrays. In these cases, we observe similar values for $\Gamma_{nr}$ and $K$ as expected from the parameters of the uncoupled resonators. The radiation loss rate associated to the $SRR$ mode is here $\widetilde{\Gamma}_{r} \approx (\Gamma_{r1}+\Gamma_{r2})/2$. The factor $1/2$ arises from the fact that the unit array cell of the combined array is actually double that of the reference array (see Fig. \ref{figII} and the insets of Figs.~\ref{figIIIi} and ~\ref{figIIIii}), and therefore the radiation loss is halved. 
For the $F=0.5$ (Fig. \ref{figIIIii}\textbf{c}) we have fitted the narrow spectral feature and its immediate high energy background; we observe good agreement for a typical radiation loss $\Gamma_{r1} \approx \Gamma_{r2} \approx 40 - 48$ $meV$ which is a good extrapolation of the values found for lower filling factors. 

\begin{figure}[h]
    \centering
\includegraphics[width=0.6\textwidth]{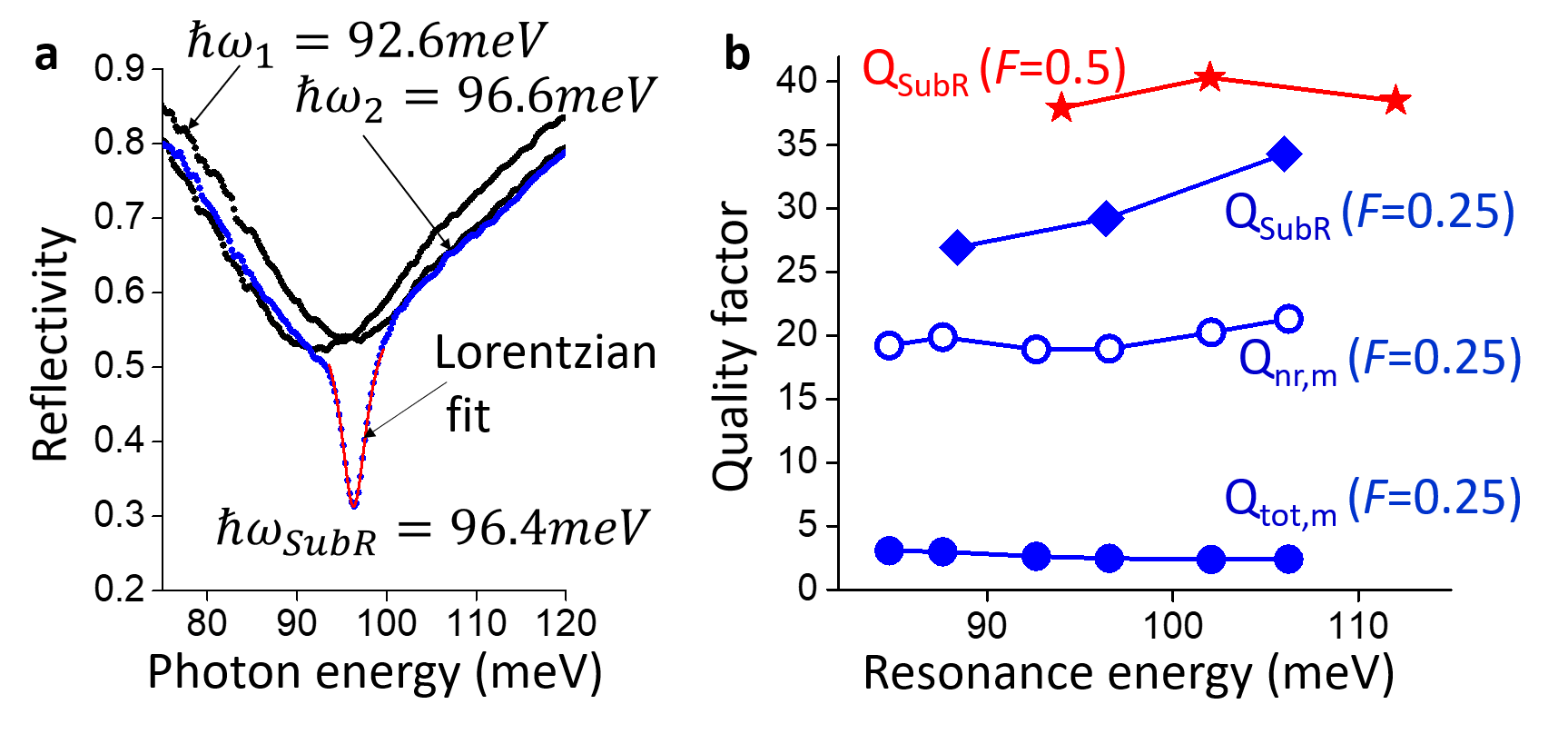}
    \caption{\textbf{a} Reflectivity spectra for the $S_1=2.1$$\mu$m $S_2=2.2$$\mu$m with $F=0.25$ (Fig.~\ref{figIIIi}\textbf{b}) and the combined $S_1+S_2$ resonance (Fig.~\ref{figIIIii}\textbf{b}). \textbf{b} Total quality factors $Q_{tot,m}$  (full dots) and non-radiative quality factors $Q_{nr,m}$ (open dots) for the individual modes of the $F=0.25$ arrays (Fig.~\ref{figIIIi}\textbf{b}). The diamonds are the quality factors $Q_{SubR}$ for the $SubR$ resonances of the $F=0.25$ arrays, and the stars are the values for the $F=0.5$, all extracted from Lorentzian fits as shown in \textbf{a}.}
    \label{figIVbis}
\end{figure}

To underline the link between the data and our model, in Fig.~\ref{figIVbis}\textbf{a} we show the reflectivity curves for the $S_1=2.1$ $\mu$m and $S_2=2.2$ $\mu$m modes from Fig.~\ref{figIIIi}\textbf{b} together with the combined spectrum from Fig. \ref{figIIIii}\textbf{b}. The similarity with Fig.~\ref{figI}\textbf{d} is obvious, and we can appreciate that the $SubR$ mode is much narrower than the original modes $\omega_1$ and $\omega_2$. To further quantify this effect from the data, we have fitted the $SubR$ resonances from Figs. \ref{figIIIii}\textbf{b,c} with an inverted Lorentzian functions $L(\omega) = B - A_{SubR}/(1 + (\omega - \omega_{SubR})^2/4\Gamma_{SubR}^2)$ where the constant $B$ takes into account the background of the $SRR$ mode.  In Fig.~\ref{figIVbis}\textbf{b} we plot the corresponding quality factors $Q_{SubR} = \omega_{SubR}/\Gamma_{SubR}$, in comparison with the total quality factors $Q_{tot,m} = \omega_{m}/(\Gamma_{rm}+\Gamma_{nrm})$ for the individual modes in \ref{figIIIi}\textbf{b} as well as their non-radiation loss factors $Q_{nr,m} = \omega_m/\Gamma_{nrm}$. Remarkably, we observe that typical values of $Q_{SubR}$ are almost tenfold higher than the individual quality factors $Q_{tot,m}$, and of the same order of magnitude as $Q_{nr,m}$. The ability to engineer such sharp resonances is the essence of the BIC phenomena \cite{PhysRevLett_2008, Bezus_2018, Hsu2016BIC,Koshelev2019BIC}. A shown in \textcolor{blue}{Supporting Information}, our model allows further studies of this phenomena as a function of the resonance detuning $\omega_2 - \omega_1$. We also show how the amplitude $A_{SubR}$ of the $SubR$ resonance can be optimized in a new type of critical coupling condition, where all incident energy can be funneled in the $SubR$ mode, even if the uncoupled resonances are far from this regime.  

\section{Conclusion}
We have demonstrated that the radiation continuum can be used to tailor BIC-type of resonances. This mechanism can be used to engineer couplings between various modes of the same structure, or modes from completely different resonators. An example is provided in \textcolor{blue}{Supporting Information}, where we show numerically  the coupling between patch cavities and meta-material resonators \cite{Pisani_2025}. Another appealing possibility is to couple patch antennas with all dielectric  \cite{Rybin2024Metaphotonics} or superconducting  resonators \cite{Chen_2010} that feature very low non-radiative loss. This type of engineering is supported by a simple analytical model using as an input the characteristics of the individual resonators. This approach opens new degrees of freedom for nanophotonics, and can be useful for various resonator-coupled emitters \cite{Urquizo_2021}, detectors \cite{Palaferri2018} and nonlinear devices \cite{Matthia2022}. 

\hspace{1cm}

\textbf{{Notes}}: 

The authors declare no competing financial interest.

\textbf{{Acknowledgment}} We acknowledge precious help from Gr\'{e}goire BEAUDOIN and Konstantinos PANTZAS as well as technical support from the clean room consortium Paris Centre. We acknowledge financial support from the European Research Council through grant ERC-COG-863487 UNIQUE, the French National Research Agency through grants ANR PRCI GERALD, ANR PRME HyQD100, as well as support from the French Renatech network.

\textbf{{Supplemental document}} \textcolor{blue}{Supporting Information} The Supporting information contains full details of the analytical model, full list of fitting parameters, description of sample fabrication, additional data with cavity thickness $H=150$ nm and $H=300$ nm,  as well as additional data and numerical modelling. 

\printbibliography

\end{document}